# A Modulator-Free Quantum Key Distribution Transmitter Chip

**Taofiq K. Paraïso[1],\***, **Innocenzo De Marco[1,2]**, **Thomas Roger[1]**, **Davide G. Marangon[1]**, **James F. Dynes[1]**,
**Marco Lucamarini[1]**, **Zhiliang Yuan[1]** and **Andrew J. Shields[1]**

[1] Toshiba Research Europe Ltd., Cambridge, UK.
[2] School of Electronic and Electrical Engineering, University of Leeds, Leeds, LS2 9JT, UK

**Quantum key distribution (QKD) has convincingly been proven compatible with real life applications. Its wide-scale deployment in optical networks will benefit from an optical platform that allows miniature devices capable of encoding the necessarily complex signals at high rates and with low power consumption. While photonic integration is the ideal route toward miniaturisation, an efficient route to high-speed encoding of the quantum phase states on chip is still missing. Consequently, current devices rely on bulky and high power demanding phase modulation elements which hinder the sought-after scalability and energy efficiency. Here we exploit a novel approach to high-speed phase encoding and demonstrate a compact, scalable and power efficient integrated quantum transmitter. We encode cryptographic keys on-demand in high repetition rate pulse streams using injection-locking with deterministic phase control at the seed laser. We demonstrate record secure-key-rates under multi-protocol operation. Our modulator-free transmitters enable the development of high-bit rate quantum communications devices, which will be essential for the practical integration of quantum key distribution in high connectivity networks.**

## Introduction

Information secrecy is an important challenge of modern society. Quantum cryptography[1], which aims at providing information theoretic security, is anticipated to be a major ingredient of future communication networks[2]. The maturity and potential of this technology is illustrated by numerous achievements such as satellite to ground[3] QKD, few nodes quantum access networks[4], long distance links[5] and novel high secret key capacity protocols[6]. To translate these notable successes into effective adoption of the technology, high bandwidth devices compatible with large-scale deployment are yet to be developed[7].

The bandwidth of quantum transmitters can be increased by multiplexing a large number of channels but this is in-scalable with bulk optics[8]. Photonic integrated circuits, which combine multiple optical components onto a small semiconductor chip, are the best candidates to respond to this demand. Recent demonstrations of all integrated indium phosphide QKD transmitters[9] and high-speed silicon photonics QKD encoders[10–12] have shown the advantage of integration in terms of stability and miniaturization of single channels. However the difficulty in realizing quantum state encoding in compact and power efficient circuits still hinders progress towards high-density integration and therefore large-scale deployment of the QKD technology.

In QKD protocols, cryptographic keys are commonly encoded in the phase of attenuated laser pulses[1]. At the core of the quantum transmitter, the quantum state encoding engine needs to be able to encode or randomize multiple phase states with deterministic phase values or with high entropy random numbers[13]. All the existing QKD photonic integrated circuits (PICs) achieve this function on-chip using high-speed interferometric modulation. This approach, also in use in classical communications[14], requires integrating multiple large footprint electro-optic modulation components, which operate at high powers and are vulnerable to chirp, residual amplitude modulation and electrical crosstalk[15]. Moreover, because of the need of phase coherence between the pulses, gain-switching is avoided and the same technique is again utilized to generate pulses from continuous wave laser sources. An approach to on-chip phase encoding free of such modulators is highly desirable as it would increase the scaling capacity and reduce the power footprint of QKD systems at the same time.

In this work, we present a QKD transmitter chip that exploits the direct phase modulation approach recently introduced in bulk optics transmitters[16]. This approach combines gain-switching, injection locking and ultra-fast phase modulation of the seed-laser to generate chirp free pulses and to realize multi-level phase encoding without the need of high-speed modulators. The lack of non-reciprocal components such as circulators or isolators in photonic integrated platforms forbids direct conversion of the bulk optics setup into photonic integrated circuits. On-chip realisation is enabled by the pulsed operation and an appropriate balance of the powers, which allow the suppression of reciprocal seeding effects from the slave laser, otherwise detrimental to deterministic phase encoding in the master laser. Our QKD transmitters can be used for time-bin encoded protocols[17,18] as well as distributed phase reference[19,20] protocols. We achieve record secure key rates of 270 kbps and 400 kbps at 20 dB attenuation (100 km in standard single mode fibre) for the decoy state BB84 and DPS protocols, respectively. Our implementation of phase encoding will also be beneficial to advanced coherent optical communications[21]. By encoding multi-level optical modulation signals with up to 8 distinct phase states at high signal integrity and low operation voltages, we demonstrate the potentialities of our new transmitter chips beyond QKD.

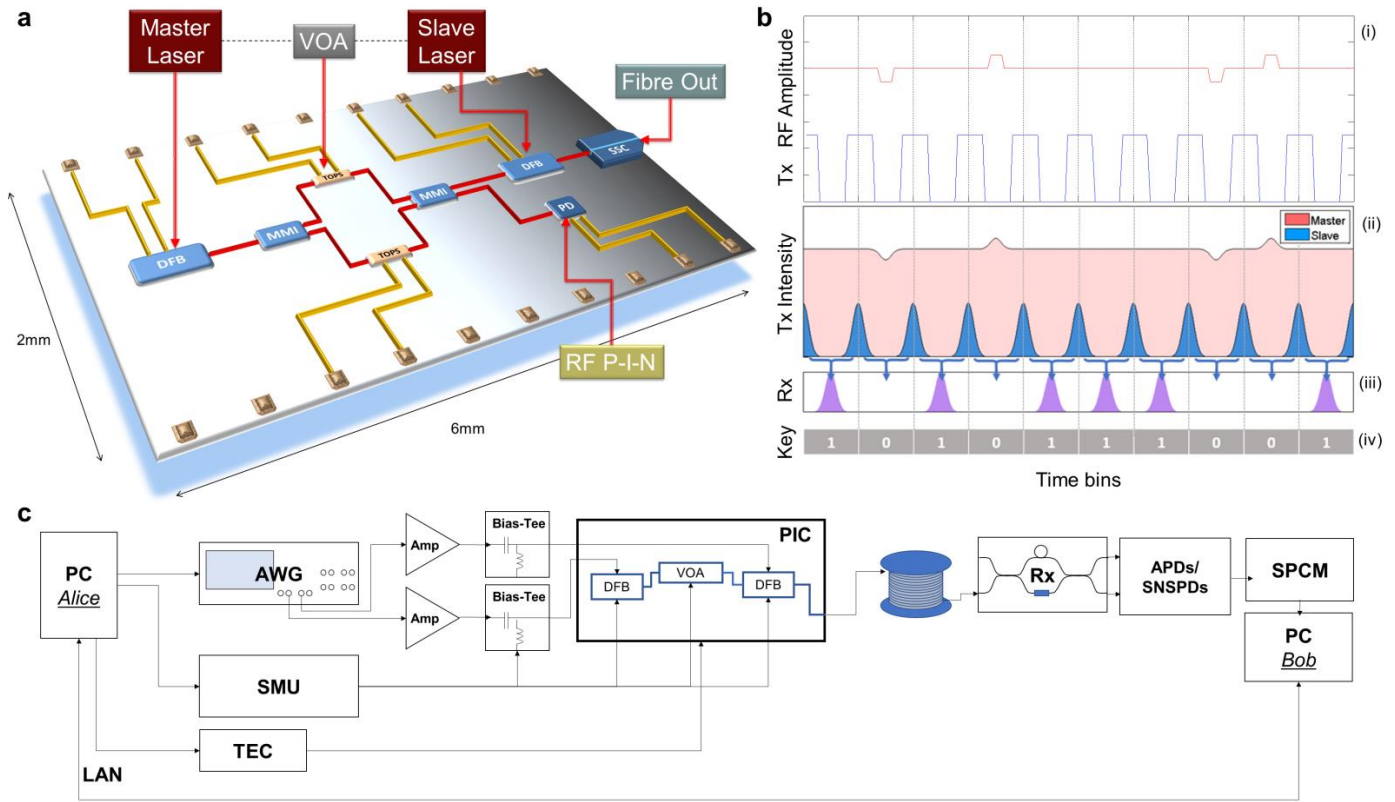

**Figure 1 | Description of the quantum transmitter chip.** (a) Simplified schematic diagram of the transmitter circuit: two DFB lasers are connected via a variable optical attenuator (VOA) to operate in injection locking. The light at the output of the Slave DFB laser is coupled to an optical fibre using a spot size converter (SSC). The VOA is a tuneable symmetric Mach-Zehnder interferometer composed of two multimode interferometers (MMI) 2x2 couplers and two DC thermo-optic phase shifters (TOPS). A high-speed photodiode (PD) is used for on-chip monitoring. (b) Principle of the phase-encoded seeding. At the transmitter (Tx), the Slave laser is gain-switched using square RF signal to generate the optical pulses (blue). Direct-phase modulation of the Master laser generates the phase-encoded optical seed (red) to be injected in the Slave cavity. The quantum key is encoded in the differential phase between the successive gain-switched pulses. At the receiver (Rx), the encoded information is retrieved by interfering consecutive pulses and a raw key is generated. (c) Experimental setup: the chip is temperature stabilized using a thermo-electric cooler (TEC). The RF signals are generated in an AWG and combined with DC signals from SMU after appropriate amplification. After propagation in a fibre-optic link, the transmitter signal is decoded in a tuneable photonic integrated delay-line interferometer followed by single photon detectors and a single photon counting module. A PC at the transmitter (Alice) controls the electronics and communicates with the receiver (Bob) via a LAN link.

## Results

**Quantum key distribution chip.** A simple schematic of the quantum transmitter chip is shown in Figure 1a. Only three main building blocks are required to generate pulse trains of phase encoded photons: two cascaded high-bandwidth distributed feedback (DFB) lasers and one optical attenuator between them. For the sake of flexibility, a thermally tuneable Mach-Zehnder interferometer is used as a variable optical attenuator. Light is coupled out of the chip into a tapered lensed fibre using a spot-size converter.

In order to drive the DFB lasers at high speed, we combine DC biases produced by precision DC current sources and radiofrequency (RF) signals from an arbitrary waveform generator (AWG). Figure 1b plots schematically the RF signals applied to the Master and Slave lasers diodes (panel i) and resulting optical signals (panel ii). To operate the Slave laser in the gain-switched regime, a 2 GHz square-wave signal is superimposed on a DC-bias below the lasing threshold, producing narrow phase-randomized pulses of 40 ps duration depicted as the blue pulses in panel ii. In our approach, the phase of each pulse from the Slave laser is locked to the variable optical phase injected from the Master laser. The Master laser phase is controlled by modulating its drive current at times synchronised between two consecutive Slave pulses, when the Slave cavity is entirely depleted. The resulting pulse train is seamlessly phase encoded: the frequency and intensity of the pulses from the Slave laser are identical but their phases differ by an offset deterministically introduced during the direct phase modulation of the Master laser. To first order, this phase shift is linearly proportional to the RF amplitude applied to the Master laser and hence differential phase shifts can be encoded arbitrarily between two consecutive pulses. The same transmitter circuit is therefore compatible with multi-level modulation signalling for classical communications, which paves the way for flexible combination of QKD and classical communications in standard infrastructures.

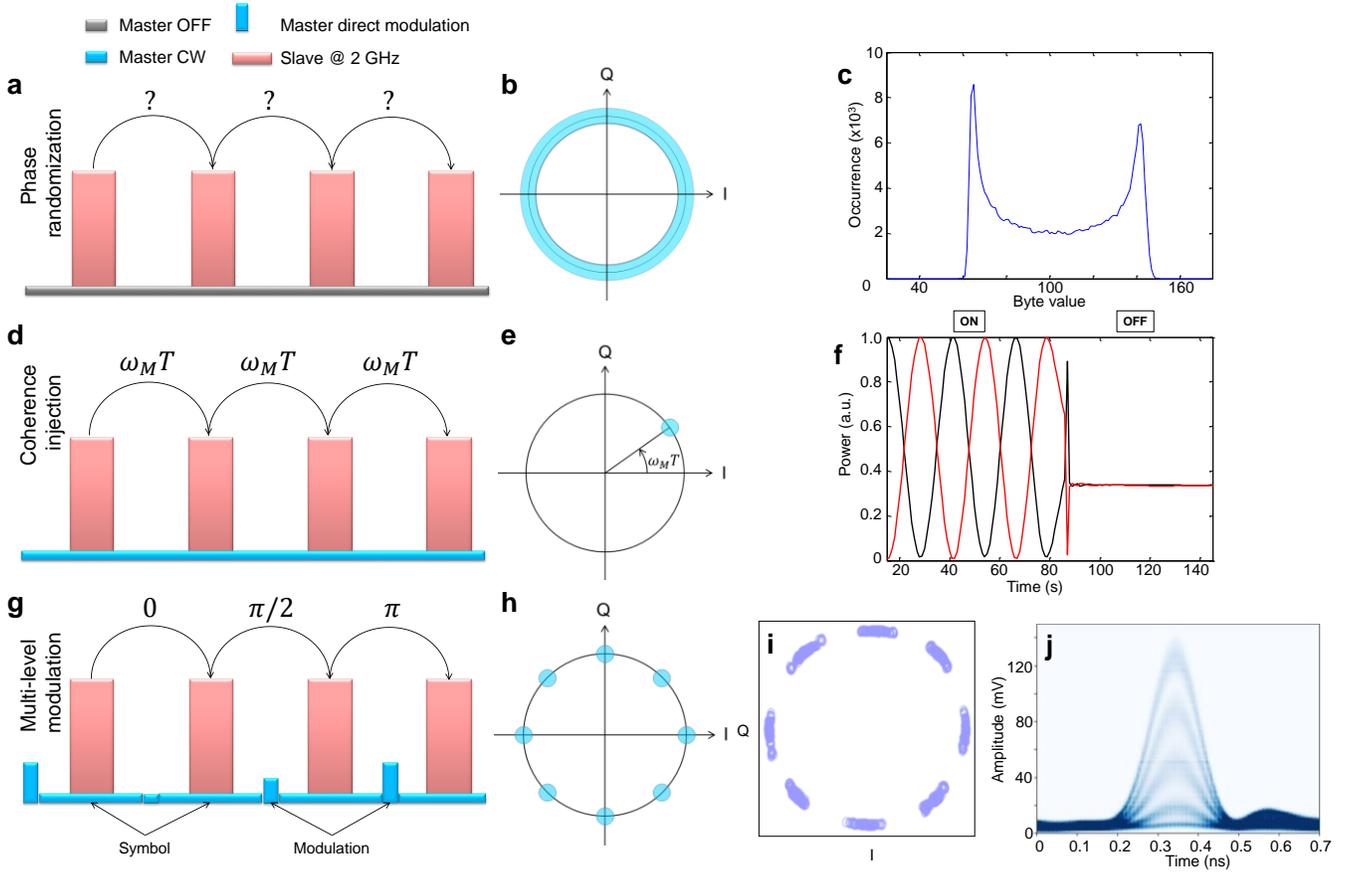

Figure 2 | On-chip deterministic phase encoding. We highlight three regimes of operation. (a-c) *No injection: phase randomization*. (a) The Slave laser is gain-switched and the Master laser is OFF. The phase relation between consecutive pulses is random because of the nature of spontaneous emission at the lasing threshold. (b) Sketch of the corresponding IQ diagram: after demodulation: we expect a ring in the complex plane with mean radius set by the pulse intensity. (c) Experimental distribution of the pulse intensities at the output of the demodulation AMZI showing the arcsine distribution characteristic of phase randomization. Quantum random numbers were extracted from this measurement. (d-f) *CW-injection: coherence transfer*. (d) The Slave laser pulses are phase locked to the CW Master laser seed. A fixed phase relation is imparted between consecutive pulses and set by the free evolution of the Master coherent wave during the repetition period T. (e) Corresponding IQ diagram expected after demodulation. (f) Interference fringes visibility measured at the output of the AMZI as the demodulation phase is swept in time. For t<87s, the injection is ON and the measured visibility is 98.3%. At t=87s, the injection laser is switched OFF and the interferences are suppressed. Note as well the decrease in mean power. (g-j) *Modulated injection: multi-level optical modulation*. (g) The Master laser modulation is synced to the edge of the symbols defined by the Slave pulse train. The phase relation between each symbol is set deterministically by the amplitude of the Master modulation. (h) Expected constellation diagram for a RZ-8DPSK with 8 modulation levels. (i-j) Experimental constellation diagram and eye diagram for a RZ-8DPSK signal after demodulation.

The phase-encoded pulse stream is decoded in a tuneable asymmetric Mach-Zehnder interferometer (AMZI) monolithically integrated on a silicon-based PIC. The time delay between the short and long arm of the AMZI is 500 ps in order to match the time delay between two consecutive pulses of the 2 GHz pulse train. The short arm features a thermal phase shifter to control the relative phase $\theta_A$ between both arms. We represent the decoded signal by the purple pulses shown in the panel iii of Figure 1b. For simplicity, we show an example in which the modulation of the master is set to introduce a $\pm \pi$ differential phase shift between pulses. For an appropriate $\theta_A$, two consecutive pulses interfere constructively or destructively depending on their differential phase. We can label constructive and destructive interference events with '1' and '0', respectively, to obtain the resultant raw key shown in panel iv of Figure 1b.. Figure 1c shows the experimental setup to test the QKD transmitter chip. A PC is used to control the laboratory equipment used to drive the transmitter (Alice). The RF waveforms from the AWG are amplified and then combined with a DC bias via bias-tees to drive the on-chip DFB lasers. The DC signals are produced using a source measure unit (SMU) and supply low noise signals to the heaters of the DFB lasers and the phase actuator of the on-chip MZI. A thermo-electric cooler (TEC), with thermistor and controller are used to maintain PIC at a stable temperature. The encoded pulses at the PIC are coupled out fibre and transmitted over the quantum channel.

The AMZI output is sent to either single photon detectors for measurement of highly attenuated QKD pulses or to high-speed photodiodes coupled to a digitizing oscilloscope for calibration using bright pulses. In the latter case, the demodulated symbols can be represented as vectors in the complex plane $re^{-i\theta}$, with $r$ the measured pulse intensity and $\theta$ the differential phase. Unambiguous recovery of $\theta$ requires two demodulation AMZIs, one measuring interferences in the $\{0,\pi\}$ basis and the other one in the $\{\pi/2, 3\pi/2\}$ basis.

**Deterministic quantum phase-encoding.** A series of measurements demonstrate the versatility of our transmitter circuit. We measure the coherence between successive pulses of the pulse train under three optical injection conditions: Master laser OFF, continuous-wave (CW) Master and directly-phase-modulated Master.

In the absence of injection, each pulse is triggered by spontaneous emission and therefore the phase relation between consecutive pulses is completely random (see Figure 2a). By representing the vectors of a large number of demodulated pulses, we obtain the expected ring shape IQ diagram sketched in Figure 2b, with constant intensity and spanning all possible phases. We exploit this phase randomization process to realize a quantum random number generator (QRNG)[22]. We send the output of the AMZI to a digital oscilloscope which records waveforms at a sampling rate of 40 GS/s and with 8-bit vertical resolution (256 values). We analyse a sequence of 1.025 million interference events and plot a histogram of the output intensities (see Figure 2c). The intensities follow the arcsine like distribution, characteristic of random phase interference (see Methods). The extracted byte autocorrelations for detection events separated up to 25ns (50 pulses) are below 5x10$^{-3}$, showing that the phase correlation between the emitted pulses are negligible. Phase randomization is an important ingredient for security in quantum communications as it effectively maps the coherent state of attenuated pulses onto photon number states. This feature prevents gain-switched laser to be used in coherent optical communication systems, unless combined with injection locking to stabilized the phase of the pulses[23].

Our second step, shown in Figure 2d-f, is to measure the effectiveness of optical injection locking using a CW seed. We bring the Master and Slave lasers into resonance using integrated heaters. Injection locking occurs at optical powers such that light injected from the Master laser overcomes spontaneous emission in the Slave laser's cavity. Jitter and chirp are then suppressed and the Slave pulses lock to the phase and wavelength of the Master. In the case of a CW Master laser described by $Pe^{i\omega_M t}$, with frequency $\omega_M$ and power $P$, the phase difference between consecutive pulses is fixed to $\Delta\varphi = \omega_M T$ where $T$ is the repetition period. Hence the corresponding IQ diagram is a vector of constant intensity and phase (figure 2e). The DFB lasers were measured to have a linewidth < 35 fm (<4.5 MHz) in CW, which is limited by the resolution of our optical spectrum analyser. This yields a coherence time in the 100 ns range which exceeds by far our repetition period $T$ = 500 ps. The coherence transfer is evidenced by resolving interference fringes at the output of the tuneable AMZI. In Figure 2f, we show that we measure interferences fringes with 98.3 % visibility when the injection is ON and that the interferences are suppressed when the injection is OFF.

Last, we realize multi-level phase encoding by combining optical injection locking and direct modulation of the Master (Figure 2g-j). The effect of direct modulation is to rotate the coherent vector in the complex plane. The 10-bit resolution of the AWG allows us to adjust this rotation angle with high precision. Hence, our transmitter can encode return-to-zero multi-level differential phase shift keying signals (RZ-MDPSK) with as many phase states $M$ as resolvable by the demodulation system. Figure 2i shows the resulting constellation diagram obtained for a 2 GHz pulse train encoded in the RZ-8DPSK format. The corresponding eye diagram showing 5 distinct levels as expected is plotted in Figure 2j.

**Multiprotocol QKD.** Thanks to the high quality of the coherence transfer on chip we can now demonstrate the record performance of our integrated quantum transmitters under different QKD protocols. We first generate random phase modulation sequences using the quantum random numbers extracted from our QRNG demonstration described above. The encoded optical pulses are then sent over an optical link before being decoded in a tuneable AMZI and measured using single photon detectors. The channel loss is set by a VOA that emulates the propagation in a standard single mode fibre, with 0.2 dB loss per km. To show the versatility of the transmitter, we realise QKD using the distributed phase shift protocol (DPS)[24] and the time bin encoded BB84 protocol[17]. These protocols require two different ways of driving the Master laser.

The results for the DPS QKD protocol are shown in Figure 3a. The Master laser is DC biased at a constant level and we encode 2 orthogonal differential phase states $\{0,\pi\}$ using the AC current modulation described above. The decoding AMZI is aligned to the correct detection basis by tuning the thermal phase shifters. We measure the decoded signal using superconducting nanowire single photon detectors (SNSPDs) with 80 % detection efficiency and dark count rate of 90 Hz. We obtain a quantum bit error rate (QBER) of 2.5 % and an asymptotic secure key rate (SKR) of 400 kb/s at 20 dB (100 km of standard single mode fibre).

For the BB84 protocol, we can encode the pulse differential phases using 4 phase states, along 2 non-orthogonal bases $X = \{0,\pi\}$ and $Z = \{\pi/2, 3\pi/2\}$. A fundamental requirement for the security of the BB84, is the randomization of the global phase between encoded pairs of pulses. Current QKD chips can only achieve phase randomization using EOPMs. Here, we use the phase randomization inherent to the gain switching by depleting the Master laser cavity between each pair of generated pulses, as schematised on Figure 3c. The Master laser is pulsed at 1 GHz with a duty cycle of 85 %. Each pulse is then directly phase modulated using an additional RF modulation. The pulses are attenuated off-chip to signal and decoy intensity levels[25,26]. We obtain a QBER of 2.2 % at 20 dB attenuation (100 km) and extract a corresponding asymptotic SKR of 270 kb/s at 20 dB (Figure 3b).

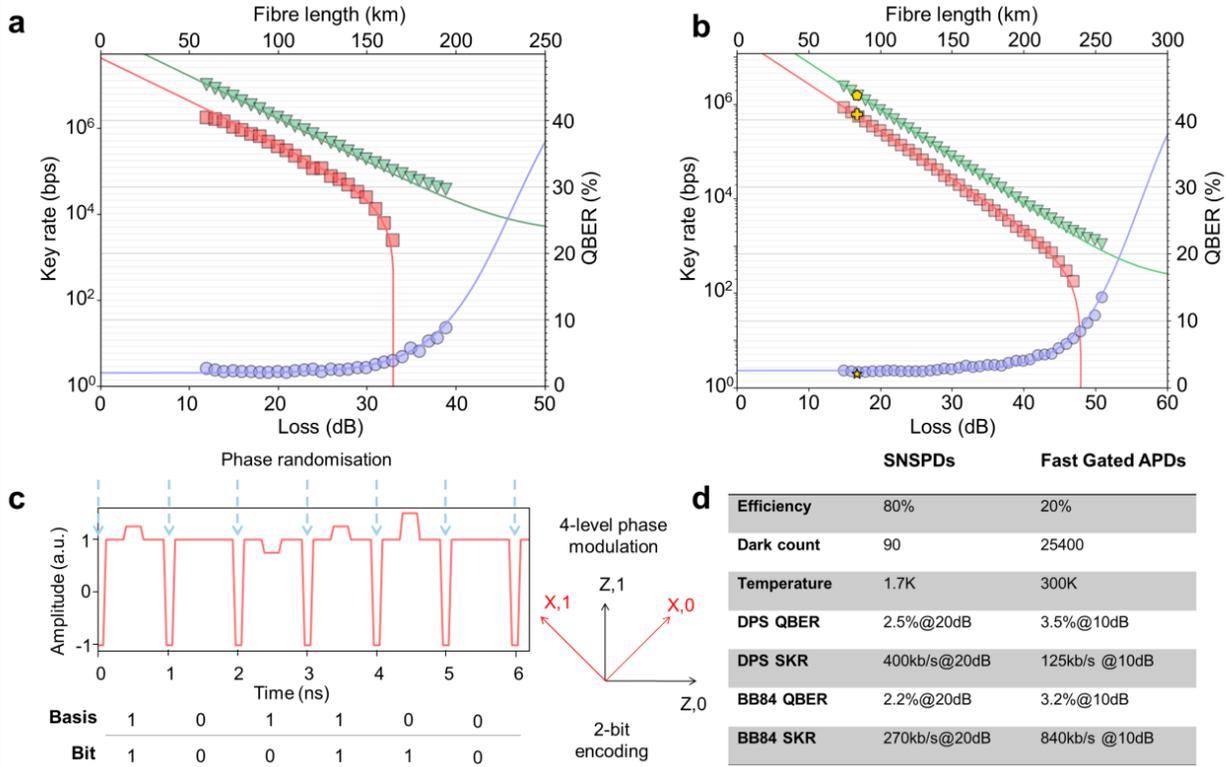

**Figure 3| Performance of the QKD transmitter chip.** Experimental QBER (circle), sifted key rate (triangles) and asymptotic secure key rates (squares) versus channel loss for (a) DPS Protocol (b) BB84 Protocol. Solid lines are analytical fits. Yellow markers are data taken over a real fibre of 75 km and 16.7 dB loss, with QBER (star), sifted key rate (pentagon) and asymptotic secure key rate (plus sign). (c) Master laser's driving electrical signals for the BB84 protocol. The complex RF signal combines multi-level optical modulation and cavity depletion for phase randomisation. During the phase randomisation, the carrier density in the Master laser is brought below the lasing threshold in order to trigger gain switching. The basis and bits of the BB84 protocol are encoded using 2-bit (4-level) direct phase modulation. (d) Comparison of the experimental QKD results using high-efficiency SNSPDs and room temperature fast-gated APDs.

We also measure the performance of our transmitter over a real fibre of 75 km deployed in our laboratory. The total loss in the fibre was 16.7 dB which is slightly higher than the 15 dB loss in an ideal standard single-mode fibre of the same length. We obtain a QBER of 2.04 %, a raw count rate of 1.54 Mc/s and an asymptotic SKR of 618 kb/s, which are in excellent agreement with the results of the emulated channel loss measurements.

To highlight the practicality of our QKD transmitter we measured QKD transmission using standard fast-gated self-differenced InGaAs avalanche photodiodes (APDs) with 18 % detection efficiency, dark count rate 25 kHz operating at room temperature[27]. A detailed comparison is shown in the table of Figure 3d. For the decoy state BB84 (DPS) protocol we measured a QBER of 3.2 % (3.5 %) and extracted an asymptotic SKR of 840 kb/s (125 kb/s) at 10 dB attenuation (50 km of standard single mode fibre), which is comparable to SKR obtained with bulk optics QKD transmitters.

We have demonstrated an on-chip QKD transmitter capable of quantum phase encoding at high clock rate and high efficiency. We achieve QKD transmission at ultra-low QBER and we establish record secure key rates for QKD chips. Our approach to complex quantum phase states encoding totally suppresses the need of on-chip modulators: our QKD transmitter circuits therefore provide the first practical and reliable route towards high-bandwidth, energy-efficient QKD optics. Multiplexing a few quantum channels will already be sufficient to overcome the most advanced bulk performance[28]. Because the protocol clock rate is set by the AC signals only, the same transmitter can be used with QKD receivers of different free spectral ranges. That is promising for Round-Robin DPS protocols[29] and for multi-node network agility.

We have also shown the compatibility of our transmitters with multi-level optical modulation formats. In coherent optical communications, these modulation formats present numerous advantages for high-bit-rate, long-haul transmission. The pulsed or return-to-zero (RZ) operation entails improved signal integrity through the reduction of fibre nonlinearity and inter-symbol interferences. This suitability for both quantum and conventional communications is a unique feature that will provide further flexibility in the combination of QKD and high-bandwidth data transmission.

**Methods**
1. **DFB lasers.** The DFB lasers were designed to emit at 1550 nm. They can be modulated at an analogue bandwidth up to 40 GHz. The resonance between the lasers was adjusted using thermal heaters embedded in the DFB laser building-block. P-I-N diodes were used to monitor the power on-chip. Light is coupled out of the chip using a spot size converter and a single mode fibre. The pulses were further attenuated off-chip using an external VOA. The RF signals driving the Master and Slave lasers are generated in an arbitrary waveform generator with 24 GS/s and 10-bit vertical resolution. The RF driving signals are synchronized and the delay between the Maser and Slave RF signals is adjusted with picosecond resolution. We use RF amplifiers to amplify the AC signals from 500mV peak-to-peak (max) up to a maximum of 3V peak-to-peak. The RF signals of the Master and Slave lasers are combined with their respective DC-bias in high-bandwidth bias-tees. The chip is mounted on a TEC controlled in continuous-current and its temperature is stabilized with mK precision.

2. **Injection power.** Simulations of our photonic circuits have shown that without attenuation, reflections at the input facet of the (undriven) Slave laser were already enough to significantly disturb the Master laser. We simulated that for 10mW output power on-chip an attenuation of 10 dB was appropriate to minimize parasitic back-reflections at the back facet of the Slave laser. Therefore our PIC features an additional 7 dB attenuator (directional coupler) between the Master and Slave lasers. We note that the MZI can be replaced with a fixed attenuator in future circuit design.

3. **QKD experiment details.** The average photon flux was set to 0.5 and 0.125 photon/pulse for the signal and decoy pulses, respectively. We consider emission probabilities of 1/16, 1/16 and 14/16 for the vacuum, decoy and signal pulses, respectively. Error correction and privacy amplification are not performed in real-time, however error-correction is taken into account in the estimated key rate through an efficiency coefficient of 90 %, which is in line with existing experiments[26].

4. **Quantum random numbers.** For the QRNG demonstration, the Slave laser is pulsed at 2 GHz with a duty-cycle of 85 % in order to eliminate artefacts from the jitter or chirp. We sample the intensities in the flat region of the gain switched pulses. We confirmed the effectivity of phase randomization and extracted quantum random numbers for repetition rates of 500 MHz, 1 GHz and 2 GHz. The arcsine distribution resulting from the interferences of the phase-randomized pulses in an AMZI with perfectly balanced input and output couplers can be derived as follows. We consider a train of gain switched pulses of equal intensity $I_{in}$ and random phases, entering one input port of the AMZI. The equation describing the intensity $I_{out}$ of the light at one output port (e.g. bar port) of the interferometer is:

$$(1) \quad I_{out}(\phi) = \frac{I_{in}}{2}[1 + \cos(\phi + \phi_0)] ,$$

where $\phi$ is the phase difference between two consecutive pulses and $\phi_0$ the relative phase between the long and short arm which can be set equal to 0 without losing generality.

The raw random numbers are associated with the successive intensities $I_{out}$ recorded at the output port of interest. To understand how the raw random numbers are distributed we derive the probability density function (PDF) of $I_{out}$, $P(I_{out})$.

Given a function $y = g(X)$ with real roots $x_1, \ldots, x_n$ such that $y = g(x_1) = \cdots = g(x_n)$, the PDF of y reads:

$$(2) \quad P(y) = \frac{P(X=x_1)}{g'(X=x_1)} + \cdots + \frac{P(X=x_n)}{g'(X=x_n)} ,$$

where $g'$ is the derivative of g with respect to X. In our case, $y = I_{out}(\phi)$, the raw random numbers, correspond to the variable $\phi$, uniformly distributed in the interval $[0,2\pi]$. Hence for a fixed $I_{in}$, Eq. (1) has exactly two roots,

$$(3) \quad \phi_{1,2} = \arccos\left(\frac{2 I_{out} - I_{in}}{I_{in}}\right) ,$$

occurring with equal probabilities $P(\phi = \phi_1) = P(\phi = \phi_2) = 1/2\pi$.

The derivative of $I_{out}$ is given by

$$(4) \quad I'_{out} = -\frac{I_{in}}{2} \sin \phi$$

evaluated on the roots

$$(5) \quad I'_{out}(\phi_1, \phi_2) = -\frac{I_{in}}{2} \sqrt{1 - \frac{(I_{in} - I_{out})^2}{I_{in}^2}}$$

Hence we have that $P(I_{out})$ takes the form of the arc-sine probability density function:

$$(6) \quad P(I_{out}) = \frac{1}{\pi \sqrt{I_{out}(I_{in} - I_{out})}} .$$

This distribution features two vertical asymptotes, one for $I_{out} = 0$ corresponding to the fully destructive interference and one for $I_{out} = I_{in}$ corresponding to the case of fully constructive interference and is an ideal representation of the distribution plotted in Figure 2c.

**Acknowledgement**

This work has been partially funded by the Innovate UK project EQUIP, as part of the UK National Quantum Technologies Programme. IDM and DGM acknowledge funding from the European Union's Horizon 2020 research and innovation programme under the Marie Skłodowska-Curie grant agreement No 675662 and 750602, respectively.

**Competing interests**

The authors declare no competing interests.

**Corresponding author**

∗ Correspondence to TKP: taofiq.paraiso@crl.toshiba.co.uk

**Author contributions**

TKP, ZY and AJS conceived the experiments. TKP designed the photonic integrated circuits. TKP, IDM and TR performed the experiments and analysed the data. DGM contributed to the QRNG experiments and analysis. JFD assisted with the installation of the single photon detectors. ML contributed to the analysis of the QKD experiments. TKP wrote the manuscript. All authors discussed the results and commented on the manuscript. ZLY and AJS supervised the project.

**Data availability**

The datasets generated during and/or analysed during the current study are available from the corresponding author on reasonable request.